# *Representations of the Canonical group, (the semi-direct product of the Unitary and Weyl-Heisenberg groups), acting as a dynamical group on noncommutative extended phase space*

Stephen G. Low
Stephen.Low@Compaq.com:
Compaq Computers,
4301 Ave. D, Austin TX, 78756

The unitary irreducible representations of the covering group of the Poincaré group P define the framework for much of particle physics on the physical Minkowski space $\mathcal{M} = \overline{\mathcal{P}}/\overline{\mathcal{L}}$, where $\mathcal{L}$ is the Lorentz group. While extraordinarily successful, it does not provide a large enough group of symmetries to encompass observed particles with a SU(3) classification. Born proposed the reciprocity principle that states physics must be invariant under the reciprocity transform that is heuristically $\{t, e, q^i, p^i\} \to \{t, e, p^i, -q^i\}$ where $\{t, e, q^i, p^i\}$ are the time, energy, position, and momentum degrees of freedom. This implies that there is reciprocally conjugate relativity principle such that the rates of change of momentum must be bounded by $b$, where $b$ is a universal constant. The appropriate group of dynamical symmetries that embodies this is the Canonical group $\mathcal{C}(1, 3) = \mathcal{U}(1, 3) \otimes_s \mathcal{H}(1, 3) = \mathcal{SU}(1, 3) \otimes_s \mathcal{O}s(1, 3)$ and in this theory the non-commuting space $\mathcal{Q} = \mathcal{C}(1, 3)/\mathcal{SU}(1, 3)$ is the physical quantum space endowed with a metric that is the second Casimir invariant of the Canonical group, $T^2 + \frac{E^2}{c^2 b^2} - \frac{Q^2}{c^2} - \frac{P^2}{b^2} + \frac{2\hbar I}{b c} (\frac{Y}{b c} - 2)$ where $\{T, E, Q_i, P_i, I, Y\}$ are the generators of the algebra of $\mathcal{O}s(1, 3) = \mathcal{U}(1) \otimes_s \mathcal{H}(1, 3)$. The idea is to study the representations of the Canonical dynamical group using Mackey's theory to determine whether the representations can encompass the spectrum of particle states. The unitary irreducible representations of the Canonical group contain a direct product term that is a representation of $\mathcal{U}(1, 3)$ that Kalman has studied as a dynamical group for hadrons. The $\mathcal{U}(1, 3)$ representations contain discrete series that may be decomposed into infinite ladders where the rungs are representations of $\mathcal{U}(3)$ (finite dimensional) or $\mathcal{C}(2)$ (with degenerate $\mathcal{U}(1) \otimes \mathcal{SU}(2)$ finite dimensional representations) corresponding to the rest or null frames.

# 1 Introduction

The unitary irreducible representations of the universal cover of the Poincaré group $\overline{\mathcal{P}}$ provides the basic framework for particle physics [1]. The Casimir invariants define the concept of mass and spin and give rise to basic equations, including the Dirac, Klein-Gordon and Maxwell equations. The quotient of $\overline{\mathcal{P}}$ with the universal cover of the Lorentz group $\overline{\mathcal{L}} = \mathcal{SL}(2, \mathbb{C})$ defines the Minkowski position-time space $\mathcal{M} = \overline{\mathcal{P}} / \overline{\mathcal{L}}$ that is regarded as the underlying *physical* space of physics. This space has the invariant $E^2 / c^2 - P^2$ where $\{E, P_i\}$, $i, j = 1, 2, 3$ are the momentum-energy degrees of freedom that may be associated with the generators of the algebra of the 4 dimensional translation group $\mathcal{T}(4)$ that is a normal subgroup of the Poincaré group.

Particle physics has proven to be much richer than can be encompassed by simply the Poincaré group. For example, $\mathcal{SU}(3)$, that is known to play a fundamental role in hadron physics, simply has no origin in the Poincaré symmetries. One of the approaches to resolve this is to increase the dimensionality of the underlying position-time *physical* space in order to give rise to groups large enough to encompass these symmetries. These dimensions are then argued to be unobservable due to their compactification into very small dimensions. In this paper, we argue that, rather then add additional position-time degrees of freedom, the physical degrees of freedom of momentum and energy are already present and must be considered in the quantum theory.

Dirac's transformation theory of quantum mechanics [2] is formulated on the non-abelian position-momentum space of the Weyl-Heisenberg group and it's associated algebra. In the non-relativistic formulation of this theory, the position and momentum degrees of freedom $\{Q_i, P_i\}$ appear to be equally fundamentally *physical* and satisfy the Heisenberg commutation relations $[Q_i, P_i] = i\hbar I$. Likewise, for time and energy, $[T, E] = -i\hbar I$.

This led Born to conjecture the notion of *reciprocity,* that the laws of physics are invariant under the reciprocal conjugation given by $\{t, e, q^i, p^i\} \to \{t, e, p^i, -q^i\}$ [3,4]. Clearly the Heisenberg commutation relations are invariant under this transform. This then led Born to the conjecture that the basic underlying *physical* space is the 8 dimensional space spanned by the degrees of freedom time-position-momentum-energy $\{T, Q_i, P_i, E\}$ with an invariant metric $-T^2 + Q^2 / c^2 + \frac{{d_{\min}}^4}{\hbar^2 c^2} (P^2 - E^2 / c^2)$ where $d_{\min}$ is a minimum length scale that Born conjectured existed. Cainiello interpreted Born's idea to be that acceleration is bounded and instead defined a maximal acceleration $a_{\max}$ as the fundamental concept [5].

The argument here is that Born's reciprocity implies that there must be a reciprocally conjugate relativity principal that leads to the rate of change of momentum (force) to be bounded by a universal constant $b$ in a manner reciprocally conjugate to the usual relativity principal that results in rates of change of position (velocity) being bounded by $c$ for particle states that have a rest frame [6][7]. Born's minimum length may be defined in terms of $b$ as $d_{\min} = \sqrt{\hbar\, c / b}$ where $b$ has the dimensions of force.

The reciprocity conjecture may be given more precise mathematical meaning by noting that the group of symmetries that leaves Born's orthogonal metric invariant is $\mathcal{O}(2, 6)$. If we also assume the symplectic structure $-E \wedge T + \delta^{i\,j} P_i \wedge Q_i$ continues to be preserved, then the group is the non-compact unitary group $\mathcal{U}(1, 3) = \mathcal{O}(2, 6) \cap \mathcal{S}p(4)$. The natural inhomogeneous group to consider is the Canonical group, the semi-direct product of the Unitary group with the Weyl-Heisenberg group $\mathcal{H}(1, 3)$, $\mathcal{C}(1, 3) = \mathcal{U}(1, 3) \otimes_s \mathcal{H}(1, 3)$ or equivalently $\mathcal{C}(1, 3) = \mathcal{SU}(1, 3) \otimes_s \mathcal{O}s(1, 3)$ where $\mathcal{O}s(1, 3)$ is the Oscillator group. This paper is essentially an exposition of the representation theory and preliminary physical consequences of this group and how it embodies Born's reciprocity principle.

The generator I is in the center of the Heisenberg algebra and is therefore a Casimir invariant, $c_1(\mathcal{C}(1, 3)) = \mathrm{I}$. The second order invariant $c_2(\mathcal{C}(1, 3)) = c_2(\mathcal{O}s(1, 3))$ is

$$c_2(\mathcal{C}(1, 3)) = -\frac{1}{2}\, \lambda_t^{-2} \left( T^2 + \frac{E^2}{c^2\, b^2} - \frac{Q^2}{c^2} - \frac{P^2}{b^2} + \frac{2\, \hbar\, I}{\mathrm{bc}} \left( \frac{Y}{\mathrm{bc}} - 2 \right) \right)$$

with $\lambda_t = \sqrt{\hbar\, / c\, b}$ and where $Y$ is the generator of the $\mathcal{U}(1)$ algebra appearing in $\mathcal{O}s(1, 3)$. This is also an invariant of both the algebras of the oscillator group and the Canonical group. Thus, the strict line element that Born conjectured must be augmented with the $\frac{2\, \hbar\, I}{\mathrm{bc}} \left( \frac{Y}{\mathrm{bc}} - 2 \right)$ term to be an invariant of the Canonical group. The appearance of this term is due to the non-abelian nature of the Weyl-Heisenberg normal subgroup.

Now, in this theory, the constant $b$ has a role that is *reciprocally conjugate* in its role to $c$ and is taken to be fundamental rather than Born's minimum length $d_{\min}$ or Cainiello's maximum acceleration $a_{\max}$. Note that as force and velocity do not commute, their respective values relative to $b$ and $c$ cannot be simultaneously observed as a measurement will yield data for one or the other depending on whether position or momentum is *diagonalized.* It is shown in [6][7] that the transformation laws, generated by the homogeneous group $\mathcal{SU}(1, 3)$ provide precisely the properties introduced here.

Born's conjecture of reciprocity directly leads to the Canonical group and the corresponding non-commuting quantum space $\mathcal{Q} = \mathcal{C}(1, 3) / \mathcal{SU}(1, 3) = \mathcal{SU}(1, 3) \otimes_s \mathcal{O}s(1, 3) / \mathcal{SU}(1, 3)$ (or the corresponding covers $\overline{\mathcal{Q}} = \overline{\mathcal{C}(1, 3)} / \overline{\mathcal{SU}(1, 3)}$). (Note that if $\mathcal{Q}$ was defined as $\mathcal{C}(1, 3) / \mathcal{U}(1, 3)$, the only natural invariant on $\mathcal{Q}$ would have been the first Casimir invariant $I$. A quadratic invariant that reduces to the usual position-time metric in the appropriate limit does not then exist).

The symmetric space $\mathcal{Q}$ is constructed from the Canonical group dynamical symmetries in a manner analogous to Minkowski space $\mathcal{M}$ being constructed from the Poincaré symmetries. However, unlike $\mathcal{M} \simeq \mathbb{R}^4$, $\mathcal{Q}$ is a non-commutative space in which the primitive points are intrinsically quantum oscillations. In the standard theory, the Poincaré group $\mathcal{P}$, from the dynamical perspective, describes how a particle state on $\mathcal{M}$ (an irreducible representation of $\mathcal{P}$) is transformed as it is rotated, boosted to a uniform rate of change of position or translated. This theory generalizes this to the dynamical Canonical group describing how a particle state on $\mathcal{Q}$ (an irreducible representation of $\mathcal{C}(1, 3)$) is transformed as it is rotated or boosted to a frame with an arbitrary rate of change of position and momentum, or translated on the non-commuting space $\mathcal{Q}$. As this intrinsically contains interacting particles (due to the non zero rates of change of momentum), one expects the dynamical symmetry to not only describe the usual particle tuplets (in the "rest" or "null" frames) but also to describe the transitions between these states when viewed from the interacting frames. That is, one expects the dynamical symmetries to be spectrum generating symmetries in the sense described by Bohm [8]. However, note that in this case, the dynamical group acts on the space $\mathcal{Q}$ itself and constructs the particle, and their associated dynamical symmetries, through the representations of this action.

The essence of the Born conjecture is that this space $\mathcal{Q}$ (and the associated Canonical group particle states) is as *physical* as the Minkowski space $\mathcal{M}$ (and its associated Poincaré group particle states). The transformations that mix the time-position degrees of freedom with the momentum-energy occur only at scales defined by $b$ which may be very large. For example, in the early universe where these scales where probably realized, we would have then notion of the position-time degrees of freedom condensing out of contracting momentum-energy degrees of freedom through these generalized $\mathcal{U}(1, 3)$ transformations.

The unitary irreducible representations of the Canonical group (or its cover) should provide a framework for particle physics that encompasses the results of the representations of the Poincaré group but are large enough to encompass symmetries known to exist but not encompassed by the Poincaré group representations. As we shall see, working out the full consequences of the reciprocity as embodied in the unitary representations of the Canonical group is a well defined, but never-the-less, formidable task. This paper outlines the required group theory and provides the general framework. A subsequent paper will examine the detailed results and the correlation with physical phenomenology.

A general method for computing semi-direct products of sufficiently well behaved Lie groups has be determined by Mackey. This theory has been used to determine the general $n, m$ dimensional Canonical group unitary irreducible representations [9][10][11].

These representations are most naturally computed on the Bargmann Hilbert space of analytic functions. The representations may be transformed to representations in which one of the subsets $\{T, Q_i\}$, $\{E, P_i\}$, $\{T, P_i\}$, $\{E, Q_i\}$ of the generators of the Heisenberg algebra are diagonal through the Segal-Bargmann transform.

## 2 Unitary Irreducible Representations of the Canonical Group

This section reviews the required group representation theory. We start with a basic properties of the Canonical group and algebra and then review the Bargmann Hilbert space of analytic functions on which the groups are represented. The Mackey representation theorems that enable the unitary dual (i.e. complete set of equivalence classes of irreducible unitary representations) for semi-direct product groups satisfying certain properties are summarized. The application to the Poincaré, Weyl-Heisenberg, Oscillator and Canonical group is then summarized. The Poincaré group is very well known, but it is presented to provide a familiar departure point for readers not familiar with Mackey theory. Computations that appear in this case also appear in the Canonical group to which we are then able to simply refer. Finally, properties certain representations of $\mathcal{U}(1,3)$ that appear in the Canonical group calculation are summarized.

### 2.1 Basic Properties of the Group and Algebra

For the partitioning $\mathcal{C}(1,3) = \mathcal{SU}(1,3) \otimes_s \mathcal{O}s(1,3)$, the group may be parameterized as $g(U, \vartheta, \omega, \iota)$ where $U \in \mathcal{SU}(1,3)$ has 15 parameters, $\theta, \iota \in \mathbb{R}$ and $\omega \in \mathbb{C}^4$. The Oscillator subgroup $\mathcal{O}s(1,3)$ has elements $g(I, \vartheta, \omega, \iota)$ and the Weyl-Heisenberg group is $g(I, 0, \omega, \iota)$. The group multiplication law is given by

$$\begin{aligned} & g(U', \vartheta', \omega', \iota')\, g(U, \vartheta, \omega, \iota) = \\ & g\left(U' U, \vartheta' + \vartheta, \omega' + e^{i\vartheta'} U' \omega, \iota' + \iota - \tfrac{i}{2}\left((\omega', U'\omega)\, e^{i\vartheta} - \overline{(\omega', U'\omega)}\; e^{-i\vartheta}\right)\right), \end{aligned} \tag{1}$$

where $(\omega, z) = \eta_{a,b}\, \overline{\omega}^a z^b$, $(U' U)^a_b = U'^a_c\, U^c_b$ and $(U\omega)^a = U^a_b\, \omega^b$. Except where otherwise noted, $a, b = 0, 1, 2, 3$, $i, j = 1, 2, 3$ and $\mathrm{diag}(\eta_{a,b}) = \{-1, 1, 1, 1\}$.

The inverse group element is

$$g(U, \vartheta, \omega, \iota)^{-1} = g(U^{-1}, -\vartheta, -U^{-1} e^{-i\vartheta} \omega, -\iota). \tag{2}$$

For the partitioning $\mathcal{C}(1,3) = \mathcal{U}(1,3) \otimes_s \mathcal{H}(1,3)$, $\vartheta$ becomes the 16[th] parameter of $U \in \mathcal{U}(1,3)$ and the group composition law may be written

$$g(U', \omega', \iota')\, g(U, \omega, \iota) = g\big(U' U, \omega' + U' \omega, \iota' + \iota - \tfrac{i}{2}\left((\omega', U' \omega) - \overline{(\omega', U' \omega)}\right)\big). \tag{3}$$

The algebra of this group is

$$[Z_{a,b}, Z_{c,d}] = \eta_{b,c} Z_{a,d} - \eta_{a,d} Z_{c,b},\ [A_a^+, A_b^-] = \eta_{a,b} I,\ [Z_{a,b}, A_c^\pm] = \mp \eta_{a,c} A_b^\pm . \tag{4}$$

Note that $\overline{A_b^+} = A_b^-$. The algebra of $\mathcal{U}(1, 3)$ factors into the direct sum of the algebra of $\mathcal{U}(1)$ and the algebra of $\mathcal{SU}(1, 3)$. The generator of the $\mathcal{U}(1)$ may be defined as $Y = \eta^{ab} Z_{ab}$ and the generators of the algebra of $\mathcal{SU}(1, 3)$ are then defined as $\hat{Z}_{ab} = Z_{ab} - \eta_{ab} Y/4$. The generators $\hat{Z}_{ab}$ satisfy the same commutation relations as given in equation (4).

The Casimir invariants of the algebra are given by [12].

$$\begin{aligned} c_1(\mathcal{C}(1, 3)) &= I, \\ c_2(\mathcal{C}(1, 3)) &= \eta^{a,b} W_{a,b}, \\ c_4(\mathcal{C}(1, 3)) &= \eta^{a,d} \eta^{b,c} W_{a,b} W_{c,d}, \\ c_6(\mathcal{C}(1, 3)) &= \eta^{a,f} \eta^{b,c} \eta^{d,e} W_{a,b} W_{c,d} W_{e,f}, \\ c_8(\mathcal{C}(1, 3)) &= \eta^{a,h} \eta^{b,c} \eta^{d,e} \eta^{f,g} W_{a,b} W_{c,d} W_{e,f} W_{g,h}, \end{aligned} \tag{5}$$

where $W_{a,b} = A_a^+ A_b^- - I Z_{a,b}$.

Note that the first Casimir invariant is simply the Casimir invariant of the Weyl-Heisenberg group and one of the two Casimir invariants of the Oscillator group, $c_1(\mathcal{C}(1, 3)) = c_1(\mathcal{H}(1, 3)) = c_1(\mathcal{O}s(1, 3))$ and the second Casimir invariant is the second Casimir invariant of the Oscillator group, $c_2(\mathcal{C}(1, 3)) = c_2(\mathcal{O}s(1, 3))$.

## 2.2 The Bargmann Hilbert Space

The group $\mathcal{C}(1, 3)$ may be represented on a 4 dimensional Bargmann space $\mathfrak{B}^4$ [14] that is the Hilbert space of entire analytic functions on $\mathbb{C}^4$ with finite inner product defined by

$$\langle f, h \rangle = \int \overline{f(z)}\, h(z)\, d\mu(z), \tag{6}$$

with $z \in \mathbb{C}^4$ and $d\mu(z) = \pi^{-4} e^{-(z,z)} d x^0 \ldots d x^3 d y^0 \ldots d y^3$ with $z^a = x^a + i y^a$.

An orthonormal basis of this space is given by $\xi_m(z) = \frac{(z^0)^{m_0}}{\sqrt{m_0!}} \ldots \frac{(z^3)^{m_3}}{\sqrt{m_3!}}$, $m_a = 0, 1, 2, \ldots$ and it follows that $\langle \xi_m, \xi_n \rangle = \delta_{m,n}$ where $\delta_{m,n} = \delta_{m_0,n_0} \ldots \delta_{m_3,n_3}$.

The Bargmann space is related to the usual Hilbert space $\mathbb{H}^4 = L^2(\mathbb{R}^4)$ of square integrable functions with inner product

$$\langle \varphi, \psi \rangle = \int \overline{\varphi(x)}\, \psi(x)\, d^4 x, \tag{7}$$

where $x \in \mathbb{R}^4$ through the Bargmann transform

$$f(z) = (\boldsymbol{B}\psi)(z) = \int B(z,x)\,\psi(x)\,d^4 x$$
$$\psi(x) = (\overline{\boldsymbol{B}}\, f)(x) = \int \overline{B(z,x)}\, f(x)\,\mathrm{d}\mu(z). \tag{8}$$

The kernel of the integral transform is given by

$$B(z,x) = \pi^{-1}\, e^{-\frac{1}{2}((z,z)+x\cdot x)+\sqrt{2}\, z\cdot x}, \tag{9}$$

where (z,z)= $\eta_{a,b}\,\bar{z}^a z^b$ , $z \in \mathbb{C}^4$ is the Hermitian inner product and $x\cdot x = \eta_{a,b}\, x^a x^b$, $x \in \mathbb{R}$ is the Lorentz inner product. In particular, the orthonormal basis $\eta_m(x)$ of $\mathbb{H}^4$ obtained by transforming the orthonormal basis vectors $\xi_m(z)$ of $\mathfrak{B}^4$ , $\eta_m(x) = (\overline{\boldsymbol{B}}\,\xi_m)(x)$ are given in terms of Hermite polynomials

$$\eta_m(x) = \eta_{m_0}(x) \ldots \eta_{m_3}(x),$$
$$\eta_{m_a}(x) = \left(2^{m_a}\, m_a!\,\sqrt{m_a}\right)^{-1/2} e^{-x\cdot x/2}\, H_{m_a}(x). \tag{10}$$

### 2.3 Mackey Representation Theory

The problem of determining the complete set of equivalence classes of unitary irreducible representations of a general class of semi-direct product groups has been solved by Mackey. The Canonical group is a special case of the $n$, $m$ dimensional Canonical group $\mathcal{C}(n,m)$ for which the Mackey theory has been applied by Wolf to obtain the full unitary dual [9].

These results are summarized in this section with explicit application of the Mackey theory to the Weyl-Heisenberg, Oscillator and Canonical groups to facilitate the physical discussion that follows.

There are two essential concepts used by Mackey to determine the unitary irreducible representations of a semi-direct product. The first is the general notion of inducing a unitary representation of a group $\mathcal{G}$ from a unitary representation of a subgroup $\mathcal{G}^\circ$. This does not require $\mathcal{G}$ to have a semi-direct product structure nor does it place any specific requirements, other than certain technical conditions, that all the cases in question satisfy, on $\mathcal{G}^\circ$ to induce representations. However, there is no guarantee that the resulting representations are irreducible or a complete set. The second of the Mackey theorems gives the construction of the specific set of subgroups $\mathcal{G}^\circ$, and the corresponding representations, in the semi-direct product case that induce the unitary dual $\hat{\mathcal{G}}$ of $\mathcal{G}$. The unitary dual is the complete set of equivalence classes of irreducible unitary representations on $\mathcal{G}$ with an appropriate Borel topology.

### 2.3.1) Mackey Induced Representation Theorem

Let $\mathcal{G}$ be a separable locally compact group with a closed subgroup $\mathcal{G}^\circ$. Let $\rho^\circ$ be unitary representations of $\mathcal{G}^\circ$ on a separable Hilbert space $\mathfrak{B}$, $\rho^\circ : \mathcal{G} \to \mathfrak{B}$. . Let $\Lambda$ be the section $\Lambda : \mathcal{G}/\mathcal{G}^\circ \to \mathcal{G}$. The representation $\rho^\circ$ on $\mathcal{G}^\circ$ induces the representation $\rho$ on $\mathcal{G}$ as:

$$(\rho(g)\, f)\,(\gamma) \;=\; \rho^\circ(\Lambda(\gamma)^{-1}\, g\, \Lambda(g^{-1}\,\gamma))\, f(g^{-1}\,\gamma) \tag{11}$$

where $g \in \mathcal{G}$, $\gamma \in \mathcal{G}/\mathcal{G}^\circ$ and $f \in L^2(\mathcal{G}/\mathcal{G}^\circ, \mathfrak{B}, \mu)$ and $\mu$ is an invariant measure on $\mathcal{G}/\mathcal{G}^\circ$.

For the cases of interest to us, the groups are very well behaved and satisfy the requisite properties and the Hilbert space is a Bargmann space of analytic functions.

### 2.3.2) Semi-Direct Group Stabilizer and Extensions Induce Unitary Dual Theorem

Now, consider the case where $\mathcal{G} = \mathcal{U} \otimes_s \mathcal{N}$ is the semi-direct product of separable locally compact groups where $\mathcal{N}$ is the normal subgroup and $\mathcal{U}$ is the homogeneous group. To emphasize this semi-direct product structure, we write an element $g$ of $\mathcal{G}$ as $g(U, n)$ where $g(U, n)\,|_{\mathcal{N}} = n \in \mathcal{N}$ and $g(U, n)\,|_{\mathcal{U}} = U \in \mathcal{U}$. We wish to find the groups $\mathcal{G}^\circ$ and the corresponding representations $\rho^\circ$ that induce through equation (11) the full unitary dual $\hat{\mathcal{G}}$. First, elements of the unitary dual $\hat{\mathcal{N}}$ of $\mathcal{N}$ are the equivalence classes of representations $[\eta]$. The actions $g \in \mathcal{G}$ on a representation $\eta$ of $\mathcal{N}$ is defined by

$$(g\,\eta)\,(n) \;=\; \eta(g^{-1}\, n\, g) \tag{12}$$

The stabilizer $\mathcal{G}^\eta$ is the subgroup of $\mathcal{G}$ that maps $\eta$ into another element of the same equivalence class, $[g\,\eta] \;=\; [\eta]$. In the case of the Abelian normal subgroup $\mathcal{N}$, equivalence is simply equality, $g\,\eta \;=\; \eta$. The stabilizer intersecting the homogeneous group is the *Little Group* $\mathcal{U}^\eta = \mathcal{G}^\eta \bigcap \mathcal{U}$ .

The subgroup $\mathcal{G}^\eta$ has a unitary dual $\hat{\mathcal{G}}^\eta$ consisting of equivalence classes of unitary irreducible representations $[\rho^\eta]$. The extension of $\eta$ is the equivalence class $[\rho^\eta]$ that, when restricted to $\mathcal{N}$, $\rho^\eta$ is a multiple of $\eta$:

$$E(\eta) \;=\; \{[\rho^\eta] \mid \rho^\eta\,|_{\mathcal{N}} =\; c\,\eta\} \text{ where } c \in \mathbb{C} \setminus \{0\} \tag{13}$$

Also, the irreducible representations $\sigma$ of the Little Group $\mathcal{U}^\eta$ may be regarded as a representation of $\mathcal{G}^\eta$ in which $\mathcal{N}$ acts trivially. The groups in question must satisfy certain topological conditions, sufficient conditions for which is that the group, normal subgroup and stabilizer group are algebraic and that $\mathcal{G}$ is analytic on $\hat{\mathcal{N}}$. The set of representations induced using the Induced Representation Theorem by the representations $\rho^\circ \;=\; \sigma \otimes \rho^\eta$ with $\mathcal{G}^\circ \;=\; \mathcal{G}^\eta$ for all $\eta \in \hat{\mathcal{N}}$ defines the complete set of irreducible representations on $\mathcal{G}$; that is the unitary dual $\hat{\mathcal{G}}$.

An important corollary of Mackey's theory is that if $\mathcal{N}$ is an abelian group, then all the equivalence classes that are elements of the dual simply have one element, one of the characters of the group. In this case, $\rho^\eta(g(U, n)) = \sigma(U) \otimes_s \eta(n)$ where $\sigma$ is an irreducible representation of $\mathcal{U}$. In general for a non-abelian normal group $\mathcal{N}$ and the expression $[g\,\eta] = [\eta]$ becomes explicitly

$$(g\,\eta)(n) = \eta(g^{-1}\,n\,g) = \chi(g)^{-1}\,\eta\,(n)\,\chi(g) \tag{14}$$

In this case, $\rho^\eta(g(U, n)) = \sigma(U) \otimes_s \chi(g(U, n))$ and $\chi(g(U, n)) = \eta(n)\,\chi(U)$ with $\chi(U) = \chi(g(U, n))\,|_{\mathcal{U}}$ a projective representation.

## 2.4 Unitary Irreducible Representations of the Poincaré group $\mathcal{P}$

The Poincaré group is the semidirect product $\mathcal{P} = \mathcal{SO}(1, 3) \otimes_s \mathcal{T}(4)$ with $\mathcal{T}(4) \simeq \mathbb{R}^4$. (We consider here the Poincaré group itself and not its universal cover $\overline{\mathcal{P}}$ but a similar analysis applies.) The group composition law is $g(L', x')\,g(L, x) = g(L'\,L, x' + L'\,x)$ where $L \in \mathcal{K} = \mathcal{SO}(1, 3)$ and $x \in \mathcal{N} = \mathcal{T}(4)$. The unitary irreducible representations of $\mathbb{R}^4$ are simply the characters $\rho_k(g(0, x)) = e^{i\,k\cdot x}$ , $k \in \mathbb{R}^4$, and the unitary dual $\hat{\mathcal{T}}$ is naturally isomorphic to the vector space dual of $\mathbb{R}^4$. The action of elements $L$ on the dual $\hat{\mathcal{T}}$ of $\mathcal{T}(4)$ is $g(L, 0)\,\rho_k = \rho_{L^{-1}\,k}$. Thus, while each of the $k_a \in \mathbb{R}$ is an invariant of the normal group $\mathcal{N}$, they are not invariants of the full Poincaré group. However, the Minkowski inner product $k\cdot k = \mu^2$ is invariant as it is a Casimir invariant of the full group. Therefore the representation theory divides into the classes of orbits based on the condition $L\,k = k$.

$$\begin{aligned} &\mathcal{O}_+ = \{\rho_k \,|\, k\cdot k = -\mu_+^2\} \text{ with } \mu_+^2 \in \mathbb{R} > 0, \\ &\mathcal{O}_- = \{\rho_k \,|\, k\cdot k = \mu_-^2\} \text{ with } \mu_-^2 \in \mathbb{R} > 0, \\ &\mathcal{O}_0 = \{\rho_k \,|\, k\cdot k = 0\}, \\ &\mathcal{O}^0 = \{\rho_0 \,|\, k = 0\}\ . \end{aligned} \tag{15}$$

For $\mathcal{O}_+$ , the stabilizer is $\mathcal{G}^\rho = \mathcal{SO}(3) \otimes_s \mathcal{T}(4)$. The cosets $\gamma_Q \in \mathcal{G}/\mathcal{G}^\rho$ are labeled by the elements $Q \in \mathcal{SO}(1, 3) \setminus \mathcal{SO}(3)$, that is the *pure* Lorentz transformations.The representations $\sigma_s(R)$, where $R \in \mathcal{SO}(3)$ and $s \in \mathbb{N}$ ($s$ is half integral for the cover $\mathcal{SU}(2) = \overline{\mathcal{SO}(3)}$). The action of the group on the cosets is $g(L, x)\,\gamma_{Q'} = \gamma_{Q\,Q'}$ and the section is $\Lambda(\gamma_Q) = g(Q, 0)$. Therefore,

$$\begin{aligned} &\Lambda(\gamma_{Q'})^{-1}\,g(L, x)\,\Lambda(\gamma_{Q^{-1}\,Q'}) = g(Q'^{-1}, 0)\,g(L, x)\ g(Q^{-1}\,Q'\,, 0) \\ &= g(Q'^{-1}\,L\,Q^{-1}\,Q',\ Q'^{-1}\,x) = g(R^\circ, Q'^{-1}\,x), \end{aligned} \tag{16}$$

where the fact that every Lorentz transformation may be written as a pure Lorentz transformation and a rotation, $L = Q\,R$ has been used to write $R^\circ = (Q^{-1}\,Q')^{-1}\,R\,Q^{-1}\,Q'$. Substituting into the induced representation theorem (11) yields

$$(\varrho_{\mu,s}(g(L, x))\,f)\,(\gamma_{Q'}) = \sigma_s\,(R^\circ)\,\rho_k(Q'^{-1}\,x)\,f\,(\gamma_{Q^{-1}\,Q'}) = \sigma_s\,(R^\circ)\,e^{i Q'\,k\cdot x}\,f\,(\gamma_{Q^{-1}\,Q'}) \tag{17}$$

There is a one-to-one bijection between pure Lorentz transformations and points $k$ with $k \cdot k > 0$. That is, for every $k, k'$ there is a unique $Q_k{}'$ such that $k' = Q_k{}' k$. Then, there is a one-one bijection between $\gamma_{Q'}$ and $k'$ giving the result

$$(\varrho_{\mu,s}(g(L, x)) f)(k') = \sigma_s(R^{\circ}) e^{i k' \cdot x} f(Q^{-1} k'). \quad (18)$$

Similar results hold for $\mathcal{O}_-$ where the stabilizer is $\mathcal{G}^{\rho} = \mathcal{SO}(1, 2) \otimes_s \mathcal{T}(4)$, with $Q \in \mathcal{SO}(1, 3) \setminus \mathcal{SO}(1, 2)$ where the representation of $\mathcal{SO}(1, 3)$ are infinite dimensional (except the trivial) and are not considered physically. The null case $\mathcal{O}_0$ where the stabilizer is $\mathcal{G}^{\rho} = \mathcal{E}(2) \otimes_s \mathcal{T}(4)$ (with $\mathcal{E}(2)$ the Euclidean group in two dimensions) and $Q \in \mathcal{SO}(1, 3) \setminus \mathcal{E}(2)$. Note that all the representations of $\mathcal{E}(2)$ are infinite dimensional except the trivial and the degenerate representations of the $\mathcal{SO}(2)$ subgroup. These finite dimensional representations are used in the physical interpretation. Finally, the degenerate case $\mathcal{O}^0$ is the representations of $\mathcal{SO}(1, 3)$ which are infinite dimensional and not considered physically.

### 2.5 Unitary Irreducible Representations of $\mathcal{H}(1, 3)$

The Weyl-Heisenberg subgroup is obtained by restricting the Canonical group law (1) to the case $U = I$, $\vartheta = 0$:

$$g(\omega', \iota') g(\omega, \iota) = g\left(\omega' + \omega, \iota' + \iota - \tfrac{i}{2}\left((\omega', \omega) - \overline{(\omega', \omega)}\right)\right). \quad (19)$$

Defining $\omega = \alpha + i\beta$ with $\alpha, \beta \in \mathbb{R}^4$, this becomes

$$g(\alpha', \beta', \iota') g(\alpha, \beta, \iota) = g\left(\alpha' + \alpha, \beta' + \beta, \iota' + \iota + (\alpha', \beta) - (\beta', \alpha)\right) \quad (20)$$

The group is therefore isomorphic to the semidirect product $\mathcal{H}(1, 3) \simeq \mathcal{K} \otimes_s \mathcal{N}$ with $\alpha \in \mathcal{K} = \mathbb{R}^4$ and $\{\beta, \iota\} \in \mathcal{N} = \mathbb{R}^5$. The abelian case of the Mackey representation theory may applies [10]. The unitary irreducible representations of $\mathbb{R}^5$ are the characters $\rho_{\upsilon,\nu}(g(0, \beta, \iota)) = e^{i(\upsilon \cdot \beta + \nu \cdot \iota)}$ where $\upsilon \in \mathbb{R}^4$, $\nu \in \mathbb{R}$. The action of the homogeneous group $\mathcal{K}$ on the dual $\hat{\mathcal{N}}$ follows from equation (12), $g(\alpha, 0, 0)\, \rho_{\upsilon,\nu} = \rho_{\upsilon+\alpha\nu, \nu}$. Thus the orbits in $\hat{\mathcal{N}}$ are:

$$\begin{aligned} \mathcal{O}_{\kappa_0} &= \{\rho_{\upsilon, \kappa_0} \mid \upsilon \in \mathbb{R}^4\} \text{ with } \kappa_0 \in \mathbb{R} \setminus \{0\}, \\ \mathcal{O}_u &= \{\rho_{u,0}\} \text{ with } u \in \mathbb{R}^4. \end{aligned} \quad (21)$$

Consider first $\mathcal{O}_{\kappa_0}$. Choosing $\rho_{0,\kappa_0}$ as the representative point in the orbit, only the identity in $\mathcal{K}$ leaves invariant $\rho_{0,\kappa_0}$ and therefore the stabilizer is $\mathcal{G}^{\rho} = \mathcal{I} \otimes_s \mathcal{N} \simeq \mathcal{N}$ and the extensions are simply the representations $\rho_{\upsilon, \kappa_0}$. The Mackey induction theorem may be applied using $\mathcal{G}/\mathcal{G}^{\rho} = \mathcal{K} \otimes_s \mathcal{N} / \mathcal{I} \otimes_s \mathcal{N} \simeq \mathcal{K}$. That is, the cosets $\gamma_\alpha \in \mathcal{G}/\mathcal{G}^{\rho}$ are labeled by the elements $\alpha$ of $\mathcal{K}$. Then $g(\alpha, \beta, \iota)\, \gamma_{\alpha'} = \gamma_{\alpha'+\alpha}$ and the section is $\Lambda(\gamma_\alpha) = g(\alpha, 0, 0)$. Substituting into the induced representation theorem (11) yields

$$
\begin{aligned}
(\eta_{\kappa_0}(g(\alpha,\,\beta,\,\iota))\,f)\,(\gamma_{\alpha'}) \;&=\; \rho_{0,\kappa_0}(\Lambda(\gamma_{\alpha'})^{-1}\,g(\alpha,\,\beta,\,\iota)\,\Lambda(\gamma_{\alpha'-\alpha}))\,f(\gamma_{\alpha'-\alpha})\\
&= \rho_{0,\kappa_0}(g(-\alpha',\,0,\,0)\,g(\alpha,\,\beta,\,\iota)\;g(\alpha'-\alpha,\,0,\,0))\,f(\gamma_{\alpha'-\alpha})\\
&= \rho_{0,\kappa_0}(g(0,\,\beta,\,\iota-\alpha'\cdot\beta))\,f(\gamma_{\alpha'-\alpha}).
\end{aligned}
\tag{22}
$$

And therefore the unitary irreducible representations of the Weyl-Heisenberg group are

$$
(\eta_{\kappa_0}(g(\alpha,\,\beta,\,\iota))\,f)\,(x) \;=\; e^{i\,\kappa_0(\iota\,-\,x\cdot\beta)}\,f(x-\alpha). \tag{23}
$$

where the identification $\mathcal{G}/\mathcal{G}^{\rho}\simeq\mathcal{K}=\mathbb{R}^4$ is used. The unitary dual that is defined by these representations is denoted $\hat{\mathcal{H}}^{\infty}$.

Using the Segal-Bargmann transform (8), this takes the form acting on the Bargmann space

$$
(\eta_{\kappa_0}(g(\omega,\,\iota))\,f)\,(z) \;= e^{\kappa_0(i\,\iota-(\omega,z-\omega/2))}\;f(z-\omega) \tag{24}
$$

where $z\in\mathbb{C}^4$ and $f(z)\in\mathfrak{B}$ is analytic.

The case $O_u$ is more straightforward. The entire group $\mathcal{K}$ leaves $\rho_{u,0}$ invariant and therefore the stabilizer is the full group $\mathcal{G}^{\rho}=\mathcal{G}$. The extensions are simply the direct product of the representations of $\mathcal{K}=\mathbb{R}^4$ with the representations $\rho_u=\;\rho_{u,0}$ of $\mathbb{R}^4$. Induction is trivial and so the representations are

$$
\eta_{u,v}(g(\alpha,\,\beta,\,\iota)) = e^{i(u\cdot\beta+v\cdot\alpha)},\;\; u,\,v\in\mathbb{R}^4, \tag{25}
$$

or defining $\omega=\alpha+i\,\beta\in\mathbb{C}$ and $w=\;v\;-\;i\,u\in\mathbb{C},\;\;\eta_w(g(\omega,\,\iota))\;=\;e^{i\,(\,w\cdot\omega\,+\,\overline{w\cdot\omega})/2}$.

The unitary dual containing these representations is denoted $\hat{\mathcal{H}}^1$ and the full unitary dual of $\mathcal{H}(1,\,3)$ is the disjoint union $\hat{\mathcal{H}}=\hat{\mathcal{H}}^{\infty}\,\bigcup\hat{\mathcal{H}}^1$.

### 2.6 Unitary Irreducible Representations of $\mathcal{O}s(1,\,3)$

The Oscillator group has the semi-direct product structure $\mathcal{G}=\mathcal{O}s(1,\,3)=\;\mathcal{U}(1)\otimes_s\mathcal{H}(1,\,3)$ [13]. In this case, the normal subgroup $\mathcal{N}=\mathcal{H}(1,\,3)$ is non-abelian for representations in $\hat{\mathcal{H}}^{\infty}$ and the homogeneous group is $\mathcal{K}=\mathcal{U}(1)$ with elements $g(\vartheta,\,0,\,0)\;=e^{i\,\vartheta}$. The action on the dual $\hat{\mathcal{N}}$ leaves $\kappa_0$ invariant as it is an eigenvalue of I that is a generates elements in the center of the Oscillator group as well as the Weyl-Heisenberg group and so the stabilizer group $\mathcal{G}^{\eta}=\mathcal{O}s(1,\,3)$. The action is

$$
\begin{aligned}
&g(\vartheta,\,0,\,0)\,\eta_{\kappa_0}(g(0,\,\omega,\,\iota))\\
&= \eta_{\kappa_0}(g(0,\,e^{i\,\vartheta}\,\omega,\,\iota))\;=\;\chi^{-1}(e^{i\theta})\,\eta_{\kappa_0}(g(0,\,\omega,\,\iota))\,\chi(e^{i\theta})
\end{aligned}
\tag{26}
$$

which is true for all $\vartheta\in\mathcal{U}(1)$. Therefore, the action elements of $\mathcal{U}(1)$ on the dual $\hat{\mathcal{H}}^{\infty}$ is $g(\vartheta,\,0,\,0)\,\eta_{\kappa_0}=\chi^{-1}(e^{i\,\vartheta})\,\eta_{\kappa_0}\,\chi(e^{i\,\vartheta})$ where $(\chi(e^{i\,\vartheta})\,f)\,(z)\;=\;f(e^{-i\,\vartheta}\,z)$. The representations of $\mathcal{U}(1)$ are $\sigma_{\kappa_1}=\;e^{i\,\kappa_1\,\vartheta}$ with $\kappa_1\in\mathbb{Z}$. The induction is trivial and therefore $\varrho_{\kappa_0,\kappa_1}=\sigma_{\kappa_1}\otimes\,\eta_{\kappa_0}\,\chi$:

$$
(\varrho_{\kappa_0,\kappa_1}(g(\vartheta,\,\omega,\,\iota))\,f)\,(z)\;=\;e^{\kappa_0(i\,\iota-(\omega,z-\omega/2))-i\,\kappa_1\,\vartheta}\;\,f(e^{-i\,\vartheta}(z-\omega)). \tag{27}
$$

For representations of $\mathcal{N}$ in $\hat{\mathcal{H}}^1$, the abelian case applies.The action of elements of $\mathcal{K}$ on $\hat{\mathcal{H}}^1$ is $g(\vartheta, \omega, \iota)\,\eta_w = \eta_{e^{-i\vartheta} w}$. The Little Group is therefore the identity and the stabilizer is $\mathcal{G}^\eta = \mathcal{I} \otimes_s \mathcal{H}(1, 3)$. The Mackey induction theorem may be applied using $\mathcal{G}/\mathcal{G}^\eta = \mathcal{U}(1) \otimes_s \mathcal{H}(1, 3)/\mathcal{I} \otimes_s \mathcal{H}(1, 3) \simeq \mathcal{U}(1)$. In this case the cosets $\gamma_\vartheta \in \mathcal{G}/\mathcal{G}^\eta$ are labeled by the elements $\vartheta$ of $\mathcal{U}(1)$. Then $g(\vartheta, \alpha, \beta, \iota)\,\gamma_{\vartheta'} = \gamma_{\vartheta'+\vartheta}$ and the section is $\Lambda(\gamma_\vartheta) = g(\vartheta, 0, 0)$. Substituting into the induced representation theorem (11) yields

$$\begin{aligned}(\varrho_w(g(\vartheta, \omega, \iota))\, f)\,(\gamma_{\vartheta'}) &= \eta_w(\Lambda(\gamma_{\vartheta'})^{-1}\, g(\vartheta, \alpha, \beta, \iota)\, \Lambda(\gamma_{\vartheta'-\vartheta}))\, f(\gamma_{\vartheta'-\vartheta}) \\ &= \eta_w(g(-\vartheta', 0, 0)\, g(\vartheta, \omega, \iota)\; g(\vartheta' - \vartheta, 0, 0))\, f(\gamma_{\vartheta'-\vartheta}) \\ &= \eta_w(g(e^{-i\vartheta'}\, \omega, \iota))\, f(\gamma_{\vartheta'-\vartheta})\end{aligned} \tag{28}$$

**2.7 Unitary Irreducible Representations of $\mathcal{C}(1, 3)$**

The form $\mathcal{C}(1, 3) = \mathcal{U}(1, 3) \otimes_s \mathcal{H}(1, 3)$ of the Canonical group is convenient to determine the representations [9]. The homogeneous group is $\mathcal{K} = \mathcal{U}(1, 3)$ and the normal group is $\mathcal{N} = \mathcal{H}(1, 3)$.

As in the case of the Oscillator group, equation (5) shows that $\kappa_0$ is an invariant of the full Canonical group $\mathcal{C}(1, 3)$. Therefore, the actions of elements of $\mathcal{U}(1, 3)$ on the dual $\hat{\mathcal{H}}^\infty$ is

$$g(U, 0, 0)\,\eta_{\kappa_0} = \chi^{-1}(U)\,\eta_{\kappa_0}\,\chi(U)\,, \tag{29}$$

where $U \in \mathcal{U}(1, 3)$ and

$$(\chi(U)\, f)\,(z) = f(U^{-1}\, z) \tag{30}$$

The stabilizer group is therefore the entire group, $\mathcal{G}^\eta = \mathcal{G} = \mathcal{C}(1, 3)$ and induction to the full group is trivial. The representations are therefore

$$\varrho_{\kappa_0,\kappa_1,\kappa_2,\kappa_3,\kappa_4}(g(U, \omega, \iota)) = \sigma_{\kappa_1,\kappa_2,\kappa_3,\kappa_4}(U) \otimes \eta_{\kappa_0}(g(\omega, \iota))\,\chi(U) \tag{31}$$

where $\sigma_{\kappa_1,\kappa_2,\kappa_3,\kappa_4} \in \hat{\mathcal{U}}$ where $\hat{\mathcal{U}}$ is the unitary dual of $\mathcal{U}(1, 3)$. As will be discussed in the following section, these representations are labelled by 4 constants corresponding to the 4 Casimir invariants of $\mathcal{U}(1, 3)$.

Note that if we had chosen to use the decomposition $\mathcal{C}(1, 3) = \mathcal{SU}(1, 3) \otimes_s \mathcal{O}s(1, 3)$, the same would have applied. In this case, from equation (5), both $\kappa_0$ and $\kappa_1$ are invariants and the action of elements of $\mathcal{SU}(1, 3)$ on the dual $\hat{\mathcal{O}s}^\infty$ is given by

$$g(U, 0, 0, 0)\,\varrho_{\kappa_0,\kappa_1} = \chi^{-1}(U)\,\varrho_{\kappa_0,\kappa_1}\,\chi(U)\,, \tag{32}$$

where now $U \in \mathcal{SU}(1, 3)$. Again, induction is trivial and the representations are

$$\varrho_{\kappa_0,\kappa_1,\kappa_2,\kappa_3,\kappa_4}(g(U, \vartheta, \omega, \iota)) = \sigma_{\kappa_2,\kappa_3,\kappa_4}(U) \otimes \varrho_{\kappa_0,\kappa_1}(g(\vartheta, \omega, \iota))\,\chi(U)\,. \tag{33}$$

It is straightforward to show that these are equivalent to the representations in equation (31).

For the case where the representations are elements of $\hat{\mathcal{H}}^1$, the representation theory uses the abelian case of the Mackey theory as $\mathcal{N} = \mathbb{C}^4$ and it is very similar to the Poincaré representation theory. The action of elements of $\mathcal{U}(1, 3)$ on the dual $\hat{\mathcal{H}}^1$ is $g(U, 0, 0)\,\eta_w = \eta_{U^{-1} w}$. The orbits are

$$\begin{aligned} \mathcal{O}_+ &= \{\eta_w \mid (w, w) = -\mu_+^2\} \text{ with } \mu_+^2 \in \mathbb{R} > 0, \\ \mathcal{O}_- &= \{\eta_w \mid (w, w) = \mu_-^2\} \text{ with } \mu_-^2 \in \mathbb{R} > 0, \\ \mathcal{O}_0 &= \{\eta_w \mid (w, w) = 0\}, \\ \mathcal{O}^0 &= \{\eta_0 \mid w = 0\} \end{aligned} \tag{34}$$

The stabilizer group for $\mathcal{O}_+$ is $\mathcal{G}^\eta = \mathcal{U}(3) \otimes_s \mathbb{C}^4$, for $\mathcal{O}_-$ it is $\mathcal{G}^\eta = \mathcal{U}(1, 2) \otimes_s \mathbb{C}^4$, for $\mathcal{O}_0$ it is $\mathcal{G}^\eta = \mathcal{C}(2) \otimes_s \mathbb{C}^4$ and for $\mathcal{O}^0$, $\mathcal{G}^\eta = \mathcal{U}(1, 3)$. The application of the induced representation then follows the Poincaré case.

For example, consider $\mathcal{O}_+$. The cosets $\gamma_Q \in \mathcal{G}/\mathcal{G}^\rho$ are labeled by *pure boosts* $Q \in \mathcal{U}(1, 3) \setminus \mathcal{U}(3)$. (The explicit form of the boosts are given in equation (52)). The action of the group on the cosets is $g(U, \omega, \iota)\,\gamma_{Q'} = \gamma_{Q\,Q'}$ where we use the fact that every element $U \in \mathcal{U}(1, 3)$ may be written as $U = Q\,R$ with $R \in \mathcal{U}(3)$ and $Q$ a unique pure boost. The section is $\Lambda(\gamma_Q) = g(Q, 0, 0)$. Then noting that

$$\begin{aligned} &\Lambda(\gamma_{Q'})^{-1}\, g(U, \omega, \iota)\, \Lambda(\gamma_{Q^{-1} Q'}) = g(Q'^{-1}, 0)\, g(U, \omega, \iota)\; g(Q^{-1} Q', 0) \\ &= g(Q'^{-1} Q\,R\,Q^{-1} Q',\; Q'^{-1} \omega, \iota) = g(R^\circ, Q'^{-1} \omega, \iota) \end{aligned} \tag{35}$$

where $R^\circ = (Q^{-1} Q')^{-1} R\, Q^{-1} Q'$. Substituting into the induced representation theorem (11) yields

$$(\varrho_{\mu,s_1,s_2,s_3}(g(U, \omega, \iota))\, f)\,(\gamma_{Q'}) = \sigma_{s_1,s_2,s_3}(R^\circ)\, \rho_k(Q'^{-1} x)\; f(\gamma_{Q^{-1} Q'}) \tag{36}$$

$\sigma_{s_1,s_2,s_3}$ are the representations of $\mathcal{U}(3)$ labeled by 3 Casimir invariants. There is a one-to-one bijection between pure boost transformations and points $z \in \mathbb{C}^4$ with $(z, z) > 0$. That is, for every $z, z'$ there is a unique $Q_z'$ such that $z' = Q_z'\, z$. Then, there is a one-one bijection between $\gamma_{Q'}$ and $z'$ giving the result

$$(\varrho_{\mu,s_1,s_2,s_3}(g(U, \omega, \iota))\, f)\,(z') = \sigma_{s_1,s_2,s_3}(R^\circ)\, \rho_{k'}(\omega, \iota)\; f(Q^{-1} z') \tag{37}$$

### 2.8 Unitary Irreducible Representations of $\mathcal{U}(1, 3)$

The unitary irreducible representations of $\mathcal{U}(n, 1)$ have been completely characterized [16][17] using methods that are a direct generalization of the methods used to characterize the irreducible representations of $\mathcal{U}(n)$ [18]. This method is based on the chain of group inclusions $\mathcal{U}(1, n) \supset \mathcal{U}(n) \supset \mathcal{U}(n-1) \ldots \supset \mathcal{U}(1)$. For the case in question $n = 3$ and the chain is $\mathcal{U}(1, 3) \supset \mathcal{U}(3) \supset \mathcal{U}(2) \supset \mathcal{U}(1)$. The representations of $\mathcal{U}(j)$ in the Gelfand basis are labeled by the $j$ integers $\{m_{1,j}, m_{2,j}, \ldots m_{j,j}\}$ with $m_{k,j} \in \mathbb{N}$ satisfying the property

$$m_{k,j} \geqslant m_{k,j-1} \geqslant m_{k+1,j} \tag{38}$$

States within the representation are given by the corresponding labels of the inclusion chain, $\{\{m_{1,j}\}, \{m_{1,j}, m_{2,j}\}, \ldots \{m_{1,j-1}, m_{2,j-1}, \ldots m_{j-1,j-1}\}\}$.

In particular the representations of $\mathcal{U}(3)$ have the invariants$\{m_{1,3}, m_{2,3}, m_{3,3}\}$ and the states within the representations are given by $\{\{m_{1,1}\}, \{m_{1,2}, m_{2,2}\}\}$. These Casimir invariants may be defined in terms of these states. The general invariants have been calculated in [19] as

$$c_n(\mathcal{U}(n)) = \sum_{i,j=1}^{n} (m_{i,n} + n - i)\, \eta_{i,j} + \theta_{i,j} \tag{39}$$

where $\theta_{i,j} = 1$ for $i < j$ and zero otherwise. In particular

$$c_2(\mathcal{U}(3)) = m_{1,3}(2 + m_{1,3}) + m_{2,3}^2 + m_{2,3}(-2 + m_{2,3}) \tag{40}$$

This may be put in a form more familiar in the physics literature (corresponding to the Cartan basis) by defining

$$n = m_{1,3} + m_{2,3} + m_{3,3}, \quad a = m_{1,3} - m_{2,3}, \quad b = m_{1,3} - m_{3,3} \tag{41}$$

and then the Casimir invariant takes the familiar form with $a$, $b$ the invariants of $\mathcal{SU}(3)$ and n the invariant of $\mathcal{U}(1)$.

$$c_2(\mathcal{U}(3)) = \frac{n^2}{3} + \frac{2}{3}\,(a^2 + b^2 - b(a-3)) \tag{42}$$

The representations of $\mathcal{U}(4)$, may be labeled by $\{\kappa_1, \kappa_2, \kappa_3, \kappa_4\} = \{m_{1,4}, m_{2,4}, m_{2,4}, m_{3,4}\}$ and the states by $\{\{m_{1,1}\}, \{m_{1,2}, m_{2,2}\}, \{m_{1,3}, m_{2,3}, m_{2,3}\}\}$. Again, the Casimir invariants may be expressed in theorem of the $\kappa_i$ and in particular,

$$c_2(\mathcal{U}(1,3)) = \kappa_1(3 + \kappa_1) + \kappa_2(1 + \kappa_2) + \kappa_3(-1 + \kappa_3) + \kappa_4(-3 + \kappa_4)\,. \tag{43}$$

This is also true for $\mathcal{U}(1,3)$ except that now the invariants $\kappa_1, \kappa_4$ are no longer necessary integers and may be complex. Also the inequalities given by (38) do not hold without modification and are generally unbounded and therefore infinite dimensional. The full representation theory shows that the constraints imposed by the group multiplication law (or locally, by the commutation relations) gives rise to seven series of representations, each with particular constraints on the $\{\kappa_1, \kappa_2, \kappa_3, \kappa_4\}$ and with appropriate inequality chains imposing conditions on the $m_{i,j}$ and $\kappa_i$. One series is the principal series and three are complementary series that parallel the Lorentz representation theory. There are however also three series of discrete representations, that do not exist in the Lorentz case, that we wish to examine more closely here.

The reason for us choosing to look only at the discrete representation is that we know that in nature particles states generally appear to be in finite dimensional representations. Clearly all the representations of this non-compact group are infinite dimensional. The discrete series however contain ladders of finite dimensional representations where the rungs are finite dimensional representations of $\mathcal{U}(3)$ that have no counterpart in the Lorentz representation theory [20]. This is very intriguing as it would give a picture of particles occupying an infinite ladder of finite dimensional representations. In fact, as will be shown in a section that follows, in the generalized rest frame, there are no transitions between the representations on different rungs and hence the representation appears to be a set of independent finite dimensional representations.

For this reason, we will review here in more detail only the discrete series $\mathcal{D}_{\pm}^{p}$, and $\mathcal{D}_{0}^{p}$, where $m_{i,j} \in \mathbb{Z}$, $\kappa_i$, $p \in \mathbb{N}$, and for $\mathcal{D}_{\pm}^{p}$ ; $\kappa_1 \geq 1$, $\kappa_4 \geq 4$.

The discrete representation cases for $\mathcal{U}(1,4)$ are [17]:

$$
\begin{aligned}
&\mathcal{D}_{+}^{1}: m_{1,3} > \kappa_1 > \kappa_4 - 4 > \kappa_2 \geq m_{2,3} \geq \kappa_3 \geqslant m_{3,3}\\
&\mathcal{D}_{+}^{2}: m_{1,3} \geqslant \kappa_2 \geq m_{2,3} > \kappa_1 + 1 > \kappa_4 - 3 > \kappa_3 \geqslant m_{3,3}\\
&\\
&\mathcal{D}_{-}^{1}: m_{1,3} \geqslant \kappa_2 > \kappa_1 > \kappa_4 - 4 > m_{2,3} \geq \kappa_3 \geqslant m_{3,3}\\
&\mathcal{D}_{-}^{2}: m_{1,3} \geqslant \kappa_2 \geq m_{2,3} \geq \kappa_3 > \kappa_1 + 1 > \kappa_4 - 3 > m_{3,3}\\
&\\
&\mathcal{D}_{0}^{3}: m_{1,3} \geq 1 + \kappa_1 \geq m_{2,3} \geq 1 + \kappa_2 \geq m_{3,3} \geq 1 + \kappa_3\\
&\mathcal{D}_{0}^{2}: m_{1,3} \geq 1 + \kappa_1 \geq m_{2,3} \geq 1 + \kappa_2;\ \kappa_4 - 1 \geq m_{3,3}\\
&\mathcal{D}_{0}^{1}: m_{1,3} \geq 1 + \kappa_1;\ \kappa_3 - 1 \geq m_{2,3} \geq \kappa_4 - 1 \geq m_{3,3}\\
&\mathcal{D}_{0}^{0}: \kappa_2 - 1 \geq m_{1,3} \geq \kappa_3 - 1 \geq m_{2,3} \geq \kappa_4 - 1 \geq m_{3,3}.
\end{aligned}
\tag{44}
$$

The representations $\mathcal{D}_0^3$ are a positive ladder of finite dimensional representations $\sigma_{n,a,b}$ of $\mathcal{U}(1) \otimes \mathcal{SU}(3)$ where $n$, $a$, $b$ are defined in (41). The representation $\eta_{\kappa_1,\kappa_2,\kappa_3,\kappa_4}$ of $\mathcal{U}(1,3)$ may be written as an infinite direct sum over $n$ of $\sigma_{n,a,b}$. For example,

$$
\begin{aligned}
&\eta_{0,0,0,0} = \oplus_{i=0}^{\infty} \sigma_{i+3,i,i},\ \eta_{1,0,0,0} = \oplus_{i=0}^{\infty} (\sigma_{i+4,i+1,i+1} \oplus \sigma_{i+5,i,i+1}),\\
&\eta_{1,1,0,0} = \oplus_{i=0}^{\infty} (\sigma_{i+6,i,i} \oplus \sigma_{i+5,i,i+1}),\ \eta_{1,1,1,0} = \oplus_{i=0}^{\infty} \sigma_{i+6,i,i},\\
&\eta_{2,0,0,0} = \oplus_{i=0}^{\infty} (\sigma_{i+5,i+2,i+2} \oplus \sigma_{i+6,i+1,i+2} \oplus \sigma_{i+7,i,i+2}) \ \ldots
\end{aligned}
\tag{45}
$$

## 3 Physical Implications

### 3.1 Physical Interpretation of the Canonical Group

The three physical constants $c$, $b$ and $\hbar$ define natural scales of time, energy, position and momentum

$$
\lambda_t = \sqrt{\hbar / b\, c}\,,\ \lambda_q = \sqrt{\hbar\, c / b}\,,\ \lambda_p = \sqrt{\hbar\, b / c}\,,\ \lambda_e = \sqrt{\hbar\, b\, c}
\tag{46}
$$

The other basic constants are dimensionless multiples of these scales, for example, $G = \alpha_G\, c^4 / b$. If an experiment was to determine that $\alpha_G = 1$, then these scales would be numerically equal to the usual Planck scales. Regardless, these scales play the role of the Planck scales in this theory.

The properties of the Canonical group may be investigated locally through the algebra. Dimensioned generators $\{T, E, Q_i, P_i\}$ may be defined as [7]

$$A_0^{\pm} = \frac{1}{\sqrt{2}}(T/\lambda_t \pm i\, E/\lambda_e), A_i^{\pm} = \frac{1}{\sqrt{2}}(Q_i/\lambda_q \pm i\, P_i/\lambda_p), \tag{47}$$

and the corresponding dimensional form of the generators of the algebra of $\mathcal{U}(1, 3)$ are

$$Z_{i,0} = N_i/b - i\, K_i/c, Z_{i,j} = M_{i,j}/b\,c - i\,\epsilon_{i,j}^k J_k, \eta^{a,b} Z_{a,b} = Y/b\,c. \tag{48}$$

An element of the unitary group $\mathcal{U}(1, 3)$ may be represented as $g(U(\beta, \gamma, \alpha, \theta, \vartheta), 0, 0) = e^Z$ and an element of the Heisenberg group as $g(0, \omega(q, p, e, t)\,\iota) = e^A$ where

$$\begin{aligned} Z &= \frac{\beta^i}{c} K_i + \frac{\gamma^i}{b} N_i + \alpha^i J_i + \frac{\theta^{i,j}}{b\,c} M_{i,j} + \frac{\vartheta}{b\,c} Y, \\ A &= \frac{1}{\lambda_t}\left(t\,T + \frac{e}{c\,b} E + \frac{q^i}{c} Q_i + \frac{p^i}{b} P_i + \frac{\iota\hbar}{\mathrm{bc}} I\right). \end{aligned} \tag{49}$$

The expanded form of the commutation relations may be straightforwardly calculated by substituting (47) and (48) into (4) [7]. Note that $\{J_i, K_i\}$ defines the usual algebra of the Lorentz subgroup of velocity boosts and that $\{J_i, N_i\}$ defines the algebra of the reciprocally conjugate Lorentz subgroup of force boosts. There are 4 Poincaré subgroups whose algebras are generated by the sets of generators $\{K_i, J_i, P_i, E\}$, $\{K_i, J_i, Q_i, T\}$, $\{N_i, J_i, Q_i, E\}$, $\{N_i, J_i, P_i, T\}$.

In terms of the algebra, the action of $\mathcal{U}(1, 3)$ on $\mathcal{H}(1, 3)$ that is defined by

$$g(0, \omega', \iota') = g(U, 0, 0)^{-1} g(0, \omega, \iota)\, g(U, 0, 0) \tag{50}$$

is simply the commutator $A' = A + [Z, A]$. This yields the transformation equations

$$\begin{aligned} T' &= T + \beta^i Q_i/c^2 + \gamma^i P_i/b^2 + \vartheta E/c^2 b^2, \\ E' &= E - \gamma^i Q_i + \beta^i P_i - \vartheta T, \\ Q_i' &= Q_i + \epsilon_{i,j}^k \alpha^j Q_k + \beta^i T - \gamma^i E/b^2 - \theta^{i,j} P_j/b^2, \\ P_i' &= P_i + \epsilon_{i,j}^k \alpha^j P_k + \beta^i E/c^2 - \gamma^i T + \theta^{i,j} Q_j/c^2. \end{aligned} \tag{51}$$

The exponential expansion of (44) may be used to compute the group action. We compute here only the *pure boost* transformations $U(\beta, \gamma, 0, 0, 0)$:

$$\begin{aligned} T' &= \cosh\zeta\, T + \frac{\sinh\zeta}{\zeta}(\beta^i Q_i/c^2 + \gamma^i P_i/b^2), \\ E' &= \cosh\zeta\, E + \frac{\sinh\zeta}{\zeta}(-\gamma^i Q_i + \beta^i P_i), \\ Q_i' &= Q_i + \left(\frac{\cosh\zeta - 1}{\zeta^2}\right)(\beta^i \beta^j/c^2 + \gamma^i \gamma^j/b^2)\, Q_j + \frac{\sinh\zeta}{\zeta}(\beta^i T - \gamma^i E/b^2), \\ P_i' &= P_i + \left(\frac{\cosh\zeta - 1}{\zeta^2}\right)(\beta^i \beta^j/c^2 + \gamma^i \gamma^j/b^2)\, P_j + \frac{\sinh\zeta}{\zeta}(\beta^i E/c^2 + \gamma^i T). \end{aligned} \tag{52}$$

where $\zeta = \sqrt{\beta^i \beta^i / c^2 + \gamma^i \gamma^i / b^2}$ . It is clear that in the limit $c, b \to \infty$ that these reduce to the usual Newtonian relations

$$\begin{aligned} T' &= T , \\ E' &= E - \gamma^i Q_i + \beta^i P_i , \\ Q_i' &= Q_i + \beta^i T , \\ P_i' &= P_i - \gamma^i T . \end{aligned} \tag{53}$$

It is clear that the rate of change of momentum (force) parameter $\gamma^i$, bounded by $b$, plays a role reciprocally conjugate to the usual rate of change of position (velocity) parameter $\beta^i$, bounded by $c$.

In this more explicit notation, substituting (47) and (48) into (5) gives the explicit form of the second Casimir invariant that was mentioned in the introduction, $c_2(\mathcal{C}(1, 3)) = c_2(\mathcal{O}s(1, 3))$ .

$$c_2(\mathcal{O}s(1, 3)) = -\frac{1}{2} \lambda_t^{-2} \Big( T^2 + \frac{E^2}{c^2 b^2} - \frac{Q^2}{c^2} - \frac{P^2}{b^2} + \frac{2 \hbar I}{b c} \left( \frac{Y}{b c} - 2 \right) \Big) \tag{54}$$

### 3.2 Physical Interpretation of the Representations of the Canonical Group

In the Poincaré theory, the particle states correspond to finite dimensional representations of the time-like and null little groups $\mathcal{SO}(3)$ and $\mathcal{SO}(2)$ depending on whether the Lorentz metric is non-zero or zero as described in equation (15) (or in the covering case, $\mathcal{SU}(2)$ and $\mathcal{U}(1)$). Poincaré transformations leave tuplets with a given Casimir invariant into themselves and in particular, there are no Poincaré transformations that take *time-like* states into *null* states.

One could think of the *time-like* states as rungs of a ladder, labeled by the Casimir invariants in representation space and that the Poincaré transformations transform these rungs into themselves with no mixing of states that are on different rungs. Likewise the null states form *null* rungs that also do not mix under the transformations.

The representations of the Canonical group has two primary cases depending on whether $\hbar = 0$ or is non-zero corresponding to the classical limit and the general quantum case.

In the $\hbar = 0$ case, the group reduces to $\mathcal{U}(1, 3) \otimes_s \mathbb{C}^4$. This case leaves the Hermitian metric invariant and has potentially physically interesting generalized time-like and null case depending on whether the Hermitian metric is non-zero or zero as described in equation (34). The the representation theory has a *time-like* case for which the *little group* is compact, $\mathcal{SO}(3)$ and $\mathcal{U}(3)$ respectively. Furthermore, the *null little groups* are $\mathcal{E}(2) = \mathcal{SO}(2)$ and $\mathcal{CE}(2) = \mathcal{U}(2) \otimes_s \mathbb{C}^2$ respectively that have compact subgroups $\mathcal{SO}(2)$ and $\mathcal{U}(2) = \mathcal{U}(1) \otimes \mathcal{SU}(2)$.

The $\hbar \neq 0$ case of the representations of the Canonical group is the general quantum case with which we are primarily concerned here. The little group in this case is the non-compact $\mathcal{U}(1, 3)$ and there is not an invariant Hermitian metric to define the time-like and null cases as above. Instead there is an additional "$\frac{2\hbar I}{b\,c}\,(\frac{Y}{b\,c} - 2)$" term in the metric representing quantum fluctuations that enables states for which the hermitian metric is zero and non-zero to be transformed into each other.

Also, in general the representations of the $\mathcal{U}(1, 3)$ can be reduced with respect to the group $\mathcal{U}(3)$ to form an infinite ladder of finite dimensional representations (45). These finite dimensional representations that these ladder rungs form are equivalent to the rest frames in this more general theory. Provided $\beta^i, \gamma^i = 0$, these rungs are transformed into each other and do not mix. However, as the general Canonical group considers non-inertial frames for which $\beta^i, \gamma^i \neq 0$, these states can now mix. Thus, the representations would be expected to generate the full particle spectra as a dynamical group. The concept of $\mathcal{U}(1, 3)$ as a dynamical group generating the particle spectra has been explored by Kalman [21], [22].

It has been noted previously that the Canonical group has 4 invariant Poincaré subgroups. These subgroups leave the Hermitian metric invariant. Consider the usual physical Poincaré subgroup. Transformations with respect to this subgroup correspond to inertial frames. In this case, the rungs of the ladder do not mix and the rungs viewed as particle states are left invariant in the sense that they only mix within the rung.

Thus,there are no true invariant time-like and null states in this representation theory and merely invariant time-like and null states under certain restricted transformations. These states, under the appropriate conditions, can be transformed amongst each other. The reason for this is contained in the *extra term* involving the $Y$ generator in the expression for the metric, the second Casimir invariant given in (54). In the degenerate classical case with $\hbar \to 0$, the term involving $Y$ vanishes and the first term is simply the Hermitian metric on abelian $\mathbb{C}^4$. Then in this case, the values of the Hermitian metric label the *time-like* and *null* states as invariant representations. In the general quantum case, only the sum of the Hermitian metric and the $Y$ term is invariant. Thus, by suitable values of $Y$ term, it can be seen that a state in a *null* frame (with zero Hermitian metric) and one in a *time-like* frame (with non-zero Hermitian metric) can have the same Casimir invariant and hence be transformed into each other. It is only through the quantum fluctuations that the $Y$ term represents that this is possible.

So far the discussion has focussed on time-like frames. It is equally possible to reduce the representations with respect to the null group, described below, that is $C(2)$. Similar arguments then apply with the null frames defining the rungs of the ladder instead of the time-like rest frames.

### 3.3 Representations of Rest and Null Frames

We can examine the representations that are the equivalent of rest and null frames in this theory. However, they will no longer in general be irreducible representations corresponding to *pure* particle states once transformed to a general frame.

Consider again the semidirect product $\mathcal{G} = \mathcal{U} \otimes_s \mathcal{N}$. The action of $\mathcal{U}$ on $\mathcal{N}$ is $n' = U^{-1}\, n\, U$. Define the subgroups $U^\circ \in \mathcal{U}^\circ \subset \mathcal{U}$ and $n^\circ \in \mathcal{N}^\circ \subset \mathcal{N}$ with the property $n^\circ = U^{\circ -1}\, n^\circ\, U^\circ$. That is, $\mathcal{G}^\circ = \mathcal{U}^\circ \otimes \mathcal{N}^\circ$ is a maximal direct product that is a subgroup of $\mathcal{G}$. Then $\mathcal{G}^\circ$ is the group $\mathcal{G}$ restricted to the generalized rest or null frame.

Before looking at the Canonical group, we briefly discuss the Poincaré group to show that, in this familiar case, this definition defines the usual time-like and null rest frames.

In this physical discussion, the groups $\mathcal{U}(3)$ and $C(2)$ are introduced as corresponding to the generalized concept of rest and null frames in the discussion. The following sections show more clearly why these groups are considered in this manner and how this relates to the Poincaré case with which we are familiar.

#### 3.3.1) Poincaré case

In the Poincaré case, the representations induced by the representations of $\mathcal{G}^\circ$ have the special property that they are isomorphic to the representations that are induced by the stabilizer group $\mathcal{G}^\eta$. This remarkable property means that the single particle states remain single particle states when viewed from a frame that has a relative velocity as one would expect in this special case. To see this, consider the time-like rest frame containing the point $x^\circ = (x^\circ, 0, 0, 0) \in \mathcal{T}(1) \simeq \mathbb{R}$. Then $\mathcal{G}^\circ = \mathcal{SO}(3) \otimes \mathcal{T}(1)$. The representations are simply $\rho_{s,\mu}\,(R, x^\circ) = \sigma_s\,(R)\, e^{i\,\mu\, x^\circ}$ where $R \in \mathcal{SO}(3)$ and $s$ is the Casimir invariant of $\mathcal{SO}(3)$ and $\mu$ is the Casimir invariant of $\mathcal{T}(1)$ [15].

Note that any Lorentz transformation can be written as $L = Q\,R$ and any point $x$ such that $x \cdot x < 0$ as $x = Q\,x^\circ$ with $Q \in \mathcal{SO}(1, 3) \backslash \mathcal{SO}(3)$. Therefore the cosets $\gamma_Q \in \mathcal{G} / \mathcal{G}^\circ$ continue to be labeled by $Q$ and $Q$ only. $\mathcal{G}^\circ$ may then be used to induce representations on $\mathcal{G}$ and the discussion continues as before. Similar arguments hold for the null case $x \cdot x = 0$ with $Q \in \mathcal{SO}(1, 3) \backslash \mathcal{E}(2)$.

#### 3.3.2) Canonical group rest and null frame

Consider then the condition $n^\circ = U^{\circ -1}\, n^\circ\, U^\circ$ for the Canonical group $C(1, 3) = \mathcal{SU}(1, 3) \otimes_s \mathcal{O}s(1, 3)$. Explicitly, this is

$$g(U^{-1}, 0, 0, 0)\, g(I, \vartheta, \omega, \iota)\, g(U, 0, 0, 0) = g(I, \vartheta, U\,\omega, \iota). \tag{55}$$

Thus the pair $U°,\ n° = (\omega°, \iota)$ are given by the condition $U°\omega° = \omega°$. This condition is satisfied by the usual 4 cases: $(\omega°, \omega°) < 0$, $\mathcal{G}° = \mathcal{SU}(3) \otimes \mathcal{O}s(1)$; $(\omega°, \omega°) = 0, \omega° \neq 0$, $\mathcal{G}° = \mathcal{C}(2) \otimes \mathcal{E}(2)$; $(\omega°, \omega°) > 0$, $\mathcal{G}° = \mathcal{SU}(1, 2) \otimes \mathcal{O}s(1)$; and the degenerate case $\omega° = 0$. We consider further only the first two cases.

### 3.3.3) Canonical group rest frame: $(\omega, \omega) < 0$ case

In this case, $\mathcal{G}° = \mathcal{SU}(3) \otimes \mathcal{O}s(1)$. The representative point in the orbit may be taken to be the *rest frame* for which only the time-energy $0^{th}$ component has non-zero values. That is, these are points in the Oscillator group for which $\{\omega^a\} = \{\omega°, 0, 0, 0\}$ that defines the one dimensional Oscillator subgroup $\mathcal{O}s(1)$ of $\mathcal{O}s(1, 3)$. The 4 generators $\{A^+{}_0, A^-{}_0, Z, I\}$ define the algebra of the one dimensional Oscillator group $\mathcal{O}s(1)$.

The representations $\sigma_{a,b}$ of $\mathcal{SU}(3)$ are given by the finite dimensional $D$ matrices[18].

The Casimir invariants restricted to $\mathcal{C}°(1, 3) = \mathcal{SU}(3) \otimes \mathcal{O}s(1)$ are

$$
\begin{aligned}
&c_1(\mathcal{G}°) = \mathrm{I} = \kappa_0 \\
&c_2(\mathcal{G}°) = A^+{}_0\, A^-{}_0 - I\ Y = c_2(\mathcal{O}s(1)) = \kappa_0\, \kappa_1, \\
&c_4(\mathcal{G}°) = (A^+{}_0\, A^-{}_0)^2 - I^2\, Y^2 - I^2\, \hat{Z}_{i,j}\, \hat{Z}_{j,i} = c_2(\mathcal{O}s(1))^2 + \kappa_0{}^2\, c_2(\mathcal{SU}(3)), \\
&c_6(\mathcal{G}°) = (A^+{}_0\, A^-{}_0)^3 - I^3\, Y^3 - I^3\, \hat{Z}_{i,j}\, \hat{Z}_{j,k}\, \hat{Z}_{k,i} = c_2(\mathcal{O}s(1))^3 + \kappa_0{}^3\, c_3(\mathcal{SU}(3))
\end{aligned}
\tag{56}
$$

where in this expression, $i, j, k = 1, 2, 3$ and the Casimir invariants for $\mathcal{SU}(3)$ are given in (41).

### 3.3.4) Canonical group null frame: $(\omega, \omega) = 0$ case

In this case, $\mathcal{G}° = \mathcal{C}(2) \otimes \mathcal{E}(2)$. In the null case, the representative point of the orbit may be taken to be $\tilde{\omega}° = \{\tilde{\omega}\circ, 0, 0, \tilde{\omega}\circ\}$. The 3 generators $\{A^{\circ\pm}, A^{\circ\pm}, Y\}$ with $A^{\circ\pm} = \frac{1}{2}\,(A_3^{\pm} - A_1^{\pm})$ satisfy the algebra

$$
[A^{\circ\pm}, Y] = \pm A^{\circ\pm}, \quad [A^{\circ+}, A^{\circ-}] = 0, \tag{57}
$$

which is the algebra of the $\mathcal{CE}(2)$ complex Euclidean group in 2 dimensions. The Casimir invariant is

$$
c_2(\mathcal{CE}(2)) = A^{\circ+}\, A^{\circ-}. \tag{58}
$$

The algebra of the group $\mathcal{U}°$ must commute with the generators $\{A^{\circ\pm}, A^{\circ\pm}, Y\}$. The maximal set of such generators are $\{C_i^{\pm},\ Z_{i,j},\ I°\}$ where $i = 1, 2$ in this section and

$$
C_i^+ = Z_{i,0} - Z_{i,3}, \quad C_i^- = Z_{0,i} - Z_{3,i}, \quad I° = Y - Z_{3,3}\ , \tag{59}
$$

where in this section $i,\ j,\ \ldots = 1, 2$. These generators satisfy the algebra

$$
[Z_{i,j}, Z_{k\,l}] = \delta_{j\,k}\, Z_{l\,i} - \delta_{i\,l}\, Z_{j\,k}, \quad [C_i^+, C_j^-] = \delta_{i\,j}\, I°, \quad [Z_{i\,j}, C_k^{\pm}] = \pm\delta_{i\,k}\, C_i^{\pm}. \tag{60}
$$

This is the algebra of the Canonical group in two dimension, $\mathcal{U}^{\circ} = \mathcal{C}(2)$. (This is analogous to the two dimensional Euclidean group appearing in the null representation case of the Poincaré group.). $\mathcal{C}(2)$ can be factored into $\mathcal{C}(2) = \mathcal{SU}(2) \otimes_s \mathcal{O}s(2)$. The corresponding parameters of the group space are $U \in \mathcal{SU}(2)$ and $g(\vartheta, \omega_i, \iota) \in \mathcal{O}s(2)$.

### 3.4 The Segal-Bargmann Transformation and the Canonical Realizations

The representations of the Canonical group have been naturally formulated on the Bargmann space of analytic functions. In this space, none of the generators $\{T, E, Q_i, P_i\}$ of the Heisenberg algebra are diagonal. The Segal-Bargmann transform may be used to transform to the Hilbert space in which a subset of these generators are diagonal. For example, we could choose to diagonalize $\{Q_i, T\}$. Then from equation (8), the transform is

$$\psi(q, t) = (\overline{\boldsymbol{B}} f)(q, t) = \int \overline{B(z, q, t)}\, f(q, t)\, \mathrm{d}\mu(z). \tag{61}$$

with $B(z, q, t) = \pi^{-1} e^{-\frac{1}{2}((z,z)-t^2+q^2)+\sqrt{2}(-z^0 t+z^i q^i)}$.

But note that we could equally well transform to spaces that diagonalize $\{P_i, T\}$ to give functions of the form $\psi(q, t)$ by using the transform kernel $B(z, p, t)$. The same also applies to the generator subsets $\{P_i, E\}$: $\psi(p, e)$ using $B(z, p, e)$ and $\{Q_i, E\}$: $\psi(q, e)$ using$B(z, q, e)$ .

It should be clear that in this intrinsically quantum theory, the representation using the Bargmann space plays a role analogous to the Hamiltonian in classical mechanics; all of the basic physical degrees of freedom $\{T, E, Q_i, P_i\}$ appear on equal, non preferential, footing. The Segal-Bargmann transform plays the role of the Legendre transform taking these to the formulations in which certain of the generators are observable and therefore diagonal. The different diagonalizations embody Born's notion of reciprocity and gives precise meaning to the heuristic expression $\{t, e, q^i, p^i\} \to \{t, e, p^i, -q^i\}$. It should be observed that even in the classical theory, where the Legendre transformation takes the Hamiltonian formulation given in terms of $\{p, q\}$ to a Lagrangian formulation in terms of $\{q, \dot{q}\}$, there is a conjugate Legendre transform that takes the Hamiltonian formulation to a Lagrangian formulation in terms of the $\{p, \dot{p}\}$. While the conjugate transform is singular for the idealized concept of a free particle, it is in all other cases valid.

# 4 Discussion and Conclusions

### 4.1 Discussion of the General Representations

Let us now summarize the picture that is emerging. From Born's very simple reciprocity principal that physical theories are invariant under the transform $\{t, e, q^i, p^i\} \to \{t, e, p^i, -q^i\}$ we are led to introduce a reciprocally conjugate relativity that bounds the rate of change of momentum by a fundamental constant $b$. This also leads us to consider the quantum non-commutative space $\mathcal{Q}= \mathcal{C}(1,3)/\mathcal{SU}(1,3)$ (or the universal covers) as the underlying *physical* space that takes on the role of the Minkowski space $\mathcal{M}= \overline{\mathcal{P}}/\overline{\mathcal{L}}$. Now, with the decomposition $\mathcal{C}(1,3) = \mathcal{SU}(1,3) \otimes_s \mathcal{O}s(1,3)$, it is clear that the space $\mathcal{Q}$ has a generalized metric $c_2(\mathcal{O}s(1,3)) = -\frac{1}{2}\lambda_t{}^{-2}\left(T^2 + \frac{E^2}{c^2 b^2} - \frac{Q^2}{c^2} - \frac{P^2}{b^2} + \frac{2\hbar I}{b\,c}\left(\frac{Y}{b\,c} - 2\right)\right)$ (54). (Note that there is no natural metric for $\mathcal{C}(1,3)/\mathcal{U}(1,3)$). Points in the space $\mathcal{Q}$ are quantum oscillations. All of the physical degrees of freedom $\{T, E, Q, P\}$ are equally physical and may be transformed into one another through the action of the homogeneous group as described in (51) and (52). It is clear that these effects will only be seen when the rate of change of momentum approaches $b$ which may be very large. These equations reduce to the expected form in the limit (53).

The idea is that the irreducible unitary representations of $\mathcal{C}(1,3)$ define the particle states of the theory on $\mathcal{Q}$ as a direct generalization of the representations of $\mathcal{P}$ giving the free particle states on $\mathcal{M}$. A key difference is that the latter considers only free particle states from uniform velocity frames with no rates of change of momentum. The more general frames in $\mathcal{C}(1,3)$ include rates of change of momentum which is indirectly related to acceleration. Thus, if we take a single particle state, we would expect it to transform into a compound state that decomposes into a sum of single particle states representing the particle interactions of the non-uniform frames.

The *Little group* of the representations of $\mathcal{C}(1,3)$ is the non-compact group $\mathcal{U}(1,3)$. It appears twice, once in the irreducible unitary representation factor and once in the projective representation factor (31). The unitary representations of this noncompact group are generally continuously infinite-dimensional. However, there are three discrete series $\mathcal{D}^p_{\pm}, \mathcal{D}^p_0$ that are comprised of infinite ladders of discrete representations (44). At least one of these series, $\mathcal{D}^3_0$, is a ladder of finite dimensional irreducible representations of $\mathcal{U}(3)$. In the rest or null frame, one is simply picking up one of the rungs in the $\mathcal{U}(1,3)$ ladder of finite dimensional representations.

The question remains as to the physical implications of the other series of representations that are infinite dimensional. In the Poincaré case, we simply discard the infinite dimensional representations as unphysical, including the $\mathcal{E}(2)$ case that shows up in the null representation as described above. It may be that particles are the ladders of finite dimensional representations and the other representations have some more field-like interpretation. As in the Poincaré case, the covering groups may be required to obtain the full particle spectrum. Clearly the representation theory of the Canonical group has a very rich texture that will take considerable effort to explore fully. While very rich in content, it is also fully defined by the group properties. Furthermore, with the introduction of the conjugate relativity principle, some of the existing assumptions about the physical interpretation of empirical data requires modification.

### 4.2 Conclusions

Born studied this idea of reciprocity over a period of more than a dozen years. This remarkably simple idea that is present in the most elementary treatments of Hamiltonian mechanics and pervades Dirac's transformation theory of quantum mechanics. By pursing it directly, one is inevitably brought to the concept of reciprocally conjugate relativity presented in this and previous papers [6], [7]. With this, one obtains the beautiful structure of the Canonical group and its representation theory. The manner in which it, along with the Segal-Bargmann transform that allows various diagonalization, completely embodies the idea of reciprocity is quite remarkable. The quantum conditions are intrinsic to the symmetry of the theory. The group has a very rich representation theory that may be adequate to encompass a significantly larger body of physics than the Poincaré representations. The manner in which the Little group, while non-compact yields infinite ladders of finite dimensional irreducible unitary representations has the potential to encompass the ever increasing array of particles is intriguing. The rest and null frames yield the groups $\mathcal{SU}(3)$, $\mathcal{SU}(2)$ and $\mathcal{U}(1)$ that appear in the standard theories.

The author is not capable of fully exploring the full spectrum of phenomena that results from this remarkably simple idea and to determine whether it correlates with the observed phenomena in this initial paper. It is the hoped that this exposition conveys the possibilities with sufficient clarity to cause further investigation of this idea.

The author wishes to thank Brian C. Hall for his very considerable help with the Mackey representation theory.